\documentclass[aps,prl,twocolumn,showpacs,superscriptaddress,groupedaddress]{revtex4}  
\usepackage{graphicx}  
\usepackage{dcolumn}   
\usepackage{bm}        
\usepackage{amssymb}   
\usepackage{amsmath}   
\usepackage{caption}
\usepackage{romannum}

\begin{document}


\title{Pattern Learning Electronic Density of States}

\author{Byung Chul Yeo} \affiliation{Computational Science Research Center, Korea Institute of Science and Technology, Seoul 02792, Korea}
\author{Donghun Kim} \affiliation{Computational Science Research Center, Korea Institute of Science and Technology, Seoul 02792, Korea}
\author{Chansoo Kim} \affiliation{Computational Science Research Center, Korea Institute of Science and Technology, Seoul 02792, Korea}
\author{Sang Soo Han} \affiliation{Computational Science Research Center, Korea Institute of Science and Technology, Seoul 02792, Korea} \email{sangsoo@kist.re.kr}


\begin{abstract}
Electronic density of states (DOS) is a key factor in condensed matter physics and material science that determines the properties of metals. First-principles density-functional theory (DFT) calculations have typically been used to obtain the DOS despite the considerable computation cost. Herein, we report a fast pattern learning method for predicting the DOS patterns of not only bulk structures but also surface structures in multi-component alloy systems by a principal component analysis. Within this framework, we use only four features to define the composition, atomic structure, and surfaces of alloys, which are the $d$-orbital occupation ratio, coordination number, mixing factor, and the inverse of miller indices. While the DFT method scales as O($\textit{N}^3$) in which $\textit{N}$ is the number of electrons in the system size, our pattern learning method can scale as O(1) regardless of $\textit{N}$. Furthermore, our method provides a pattern similarity of 91$\sim$98$\%$ compared to DFT calculations. This reveals that our learning method will be an alternative that can break the trade-off relationship between accuracy and speed that is well known in the field of electronic structure calculations.
\end{abstract}

\maketitle
\begin{center}
\textbf{\Romannum{1}. INTRODUCTION}
\end{center}

Electronic density of states (DOS) plays a tremendously important role in determining the properties of metals\cite{Martin}. Researchers in the fields of solid-state and condensed matter physics carefully diagnose density distributions of free electrons in metals to understand scientific concepts that are hidden in such density distributions (e.g., the $d$-band center theory)\cite{Norskov2009} and to develop new materials\cite{Seo2014,Ma2015}.

Quantum mechanical approaches (e.g., density functional theory) shed light on the nature of electrons in metals, and first-principles density functional theory (DFT) calculations are successful methods to develop the electronic DOS of metals. Although quantum mechanical methods provide a high accuracy, they have the disadvantage of a severe computational workload, which originates from the complexity of many-body systems\cite{Ratcliff2017}. Thus, many researchers are seeking a fast method to predict electronic structures of materials with a high accuracy\cite{Galli1996,Saad2010,Goedecker1999,Ordej1998}. 

Within quantum mechanical frameworks, their high computational cost limits the system size that can be studied. To circumvent such causality-based frameworks, an inductive method can be realized by utilizing data and statistical learning algorithms\cite{Lecun2015,Jordan2015,Carleo2017,Biamonte2017,Carrasquilla2017,Nieuwenburg2017,Snyder2012,Ghiringhelli2015}. Recently, a machine-learning approach was pursued to address different quantum mechanical problems\cite{Brockherde2017,Arsenault2014}, and in particular, to predict the electronic structures of alloys, e.g., to predict the DOS values at the Fermi level\cite{Schutt2014} or the $d$-band centers\cite{Takigawa2016}. However, to date, these attempts have been limited to the prediction of only single value, and no machine-learning technique is available for the prediction of DOS patterns that includes both the value and shape.

Herein, we propose a new perspective on the representation of DOS that has been regarded as multi-dimensional digital data from one-dimensional continuous curves. Using principal component analysis, we identified highly correlated DOS patterns for various metal systems and proposed features to determine the correlation between the DOS patterns and the atomic structures of materials in a linear subspace. We successfully reproduced the DOS patterns of alloys usually found by quantum mechanical approaches but with a zero scaling, O(1), which is independent of the number of electrons in the system. Furthermore, our method achieves a small loss of the DOS patterns compared to DFT calculations. The DOS pattern learning method can provide a breakthrough in the trade-off relationship between accuracy and speed, which is well known in the field of electronic structure calculations. Moreover, the approach is applicable for predicting DOS patterns in not only bulk structures but also of surfaces in multi-component alloy systems.

\begin{center}
\textbf{\Romannum{2}. METHODS}
\end{center}

When mapping DOS patterns from the atomic structures of alloys, there is a mathematical puzzle, i.e., the number of input material labels (e.g., compositions, crystal structures, and lattice parameters) is much smaller than the number of output DOS values at the corresponding energy levels. Accordingly, we first compressed the output information by digitizing an analog signal of the DOS in a rectangular window to one multi-dimensional vector, as shown in Fig. 1a. Next, we applied principal component analysis (PCA), an unsupervised learning technique, to reduce the high-dimensional data to a low-dimensional data set\cite{Mueller2016,Hastie}. Then, we could build a model to represent the DOS patterns.

\begin{figure*}
\includegraphics[width=0.90\textwidth]{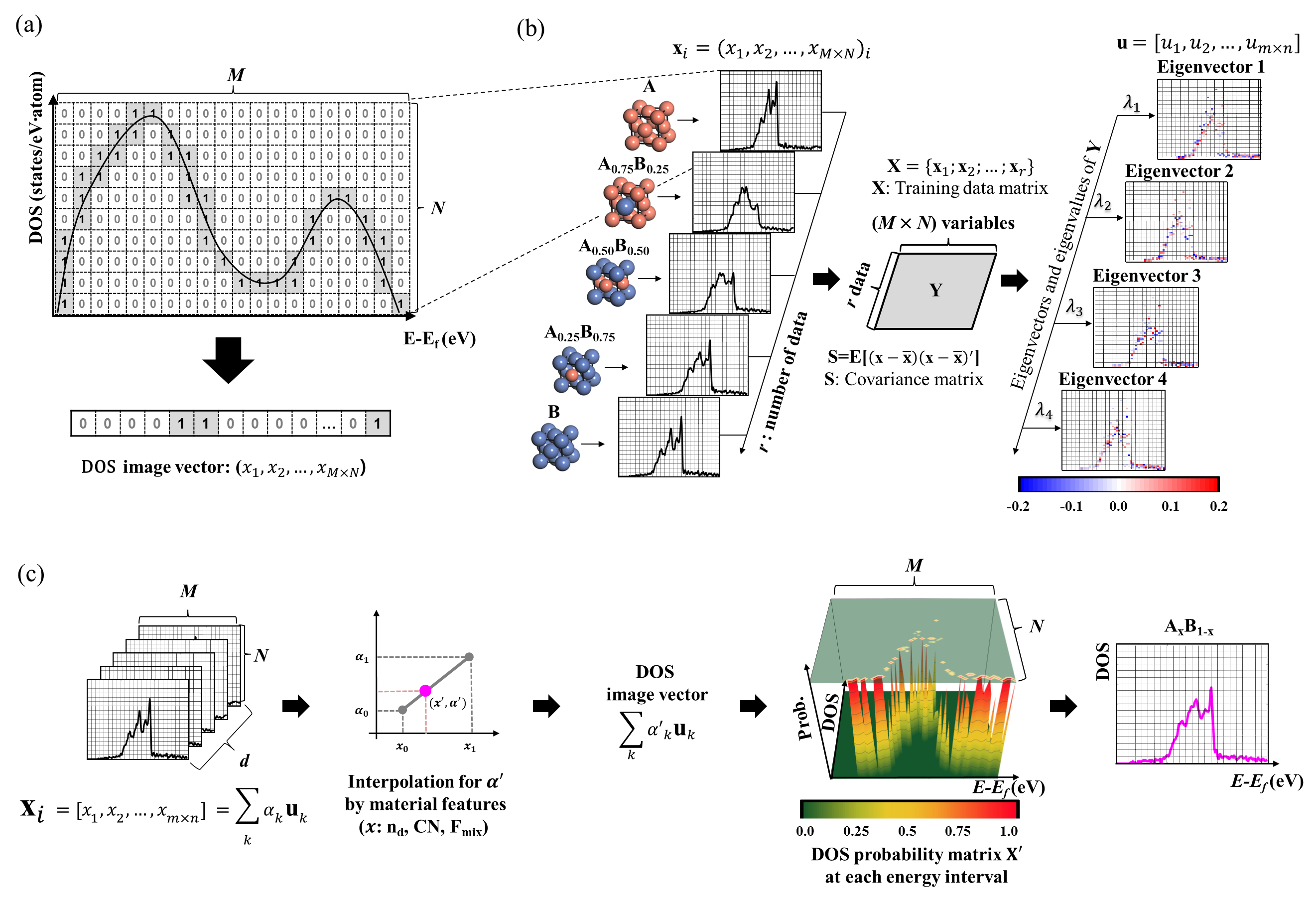}
\caption {Scheme of the pattern learning (PL) method for learning and predicting electronic DOSs. (a) Conversion of a DOS pattern from a continuous energy function in a rectangular window to a digital image vector with $M \times N$ entries. (b) Learning process of PCs of $\text{A}_x$$\text{B}_{1-x}$ alloys with their DOS patterns. $x_{i}$ is a row vector where $M$ and $N$ correspond to the grid size of the DOS window, and $\mathbf{\bar{x}}$ is the average value of the entries in the row vectors. As a training system for learning, five compositions\{A, $\text{A}_{0.75}$$\text{B}_{0.25}$, $\text{A}_{0.5}$$\text{B}_{0.5}$, $\text{A}_{0.25}$$\text{B}_{0.75}$, and B\} are considered on the left side. A covariance matrix, $\mathbf{Y}$, is constructed in the middle. PCA determine the eigenvectors, which are PCs, and eigenvalues of the training data set, which are shown on the right side. (c) The prediction process of an unknown DOS pattern for an arbitrary alloy, $\text{A}_{x}$$\text{B}_{1-x}$. The process involves several steps: 1) estimation of PCs coefficients using features, including $\text{n}_d$, CN, $\text{F}_{mix}$; 2) estimation of a new DOS image vector; 3) production and utilization of the DOS probability matrix; and 4) prediction of the DOS pattern for the test alloy, $\text{A}_x$$\text{B}_{1-x}$, using a probability matrix.}
\label{fig:FIG. 1}
\end{figure*}

\begin{center}
\textbf{A. Learning process of DOS patterns}
\end{center}

In the learning process of the DOS pattern ($\rho$), PCA was employed in which, we implemented Python code with matrices operation package \textit{NumPy}\cite{Ramani2017} for the analysis. Mathematically, this code finds the maximum variance of linearly independent eigenvectors. Prior to the analysis, DOS image vectors were digitized in a rectangular window. In our study, we considered an energy range from -10 eV to 5 eV and a DOS range from 0 to 3. We standardized the DOS image vectors of the training data by obtaining the normalized matrix $\textbf{Y}$ in which the $i$ th row ($\textbf{y}_i$) of $\textbf{Y}$ is $\mathbf{x_\textit{i} - \bar{x}}$, where $\mathbf{\bar{x}}$ is the mean of each column vector of \textbf{X}. Then, we calculated the eigenvectors, $\textbf{u}_p$ =($u_1$, $u_2$, \ldots, $u_{M \times N}$$)_p$, and the corresponding eigenvalues, $\lambda_p$, were calculated by the covariance matrix, $\mathbf{S = Y^TY}$, according to eq. (1):  

\begin{equation}
\mathbf{Su}_p = \mathbf{\lambda}_p \mathbf{u}_p
\end{equation}

Here, the eigenvectors are called \textit{principal components} (PCs), and the corresponding eigenvalues describe the data variance along the PCs. 

The original vector $\mathbf{x}$ can be reconstructed by using the following eq. (2):

\begin{equation}
\mathbf{x} \approx \sum_{p=1}^{P} (\mathbf{y^T} \mathbf{u}_p)  \mathbf{u}_p + {\sum_{p=1}^{P} (\mathbf{\bar{x}}^T  \mathbf{u}_p)  \mathbf{u}_p} = {\sum_{p=1}^{P} \alpha_p  \mathbf{u}_p} 
\end{equation}

where $P$ is the number of PCs and $p$ is their index. Thus, coefficient $\alpha_p$ of the eigenvectors can be computed by $\mathbf{y^T}  \mathbf{u}_p + \mathbf{\bar{x}^T}  \mathbf{u}_p$, and it corresponds to the coordinate values on the linear subspace that is composed of PCs.

In the learning process using the PCA, we identified the linear subspace for which the orthogonal projections of the image vector, $x$, have a maximum variance, and we learned the eigenvectors, $\mathbf{u}$, of the training systems in the linear subspace (Fig. 1b). The original image vectors can be reconstructed by $\sum_{p=1}^{P} \alpha_p \mathbf{u}_p$ .

\begin{center}
\textbf{B. Predicting process of DOS patterns}
\end{center}

During the predicting process for the DOS pattern (a new image vector, $\mathbf{x'}$) of a test alloy, as shown in Fig. 1c, we estimated the new coefficients, $\alpha'_p$, via a linear interpolation between $\alpha_p$ of the two training systems that is most similar to the test composition, where features relevant to the electron occupation and atomic configuration were considered (Supplementary Figs. S1 and S2). Using $\sum_{p=1}^{P} \alpha'_p \mathbf{u}_p$, we obtained a new image vector, $\mathbf{x'}$, and transformed from the $\mathbf{x'}$ to the DOS probability matrix, $\mathbf{X'}$, the elements of which are the probable values of each DOS levels at the given energy interval. 

To predict DOS patterns, only a single DOS value must be determined by a given energy interval. Thus, we defined the DOS probability matrix originating from the DOS image vector. The DOS image vector, $\mathbf{x'}$ =($x_1'$, $x_2'$, \ldots, $x_{M \times N}'$), calculated by the PCs and the estimated coefficients, was transformed to the DOS image matrix, $\mathbf{I'}$, with $M$ columns and $N$ rows in a grid-based rectangular window, and its size is the same as the size used in the learning process (Supplementary Figs. S5). To define the DOS probability matrix, we considered only positive entries in the $\mathbf{I'}$, and the other entries were regarded as zero. Moreover, we normalized all of the entries of the DOS levels at each energy interval. Then, we defined the DOS probability matrix, $\mathbf{X'}$, with $M$ columns and $N$ rows, as given by 

\begin{equation}
\mathbf{X'}_{m,n}=\frac{x'_{m,n}}{\sum_{n} x'_{m,n}}
\end{equation}

where $x'_{m,n}$ is the positive entry value of the column vector in $\mathbf{X'}$, and $m$ and $n$ are the matrix indices. 

To predict the DOS pattern with $\mathbf{X'}$, one should determine a single DOS value at each energy interval. Therefore, we obtained the estimated DOS, which is $\rho'$ and is given by

\begin{equation}
\rho' = \sum_{m=1}^M \rho' (E_m) = \sum_{m=1}^M \sum_{n=1}^N { \mathbf{X'}_{m,n} \cdot \rho_n (E_m)}
\end{equation}

where $E_m$ is the $m$ th energy interval, and $\rho_n$ is the $n$th DOS level.

\begin{center}
\textbf{\Romannum{3}. RESULTS}
\end{center}

\begin{center}
\textbf{A. Application into binary alloy systems}
\end{center}

To test our pattern learning (PL) method, it was first applied to a Cu-Ni system. Thermodynamically, this alloy system shows a complete solid solution, indicating that the Cu and Ni atoms in the alloys are homogeneously mixed in a face-centered cubic (fcc) structure regardless of the composition. Thus, it is expected that the DOS of the alloy system follows intrinsic electronic structures of Cu and Ni crystals and their composition can be a key feature for the representation of their DOS patterns. Therefore, we define the $d$-orbital electron occupation rate ($n_{d}$) as an alloy composition-dependent feature that represents local DOS patterns of the $d$-orbitals. Moreover, all of the pristine Cu and Ni and their alloys have a fcc crystal structure, indicating that the effect of the atomic structure on the DOS pattern is not significant. Accordingly, to predict the DOS patterns in this system, we considered only nd as a feature. After training the DOS patterns for various Cu-Ni compositions (Cu, $\text{Cu}_{0.75}$$\text{Ni}_{0.25}$, $\text{Cu}_{0.5}$$\text{Ni}_{0.5}$, $\text{Cu}_{0.25}$$\text{Ni}_{0.75}$, Ni) , we predicted the DOS of $\text{Cu}_{0.375}$$\text{Ni}_{0.625}$ as the test alloy (Fig. 2a) by considering three PCs. A comparison with the DFT results revealed that our method obtained the pattern similarity of 95\% ($\sigma$ = 0.95). However, the calculation time is less than 1 minute even on 1 core of an Intel Xeon CPU, whereas the DFT method requires approximately 2 hours on 16 cores of the CPU.

\begin{figure}[h]
\includegraphics[width=0.48\textwidth]{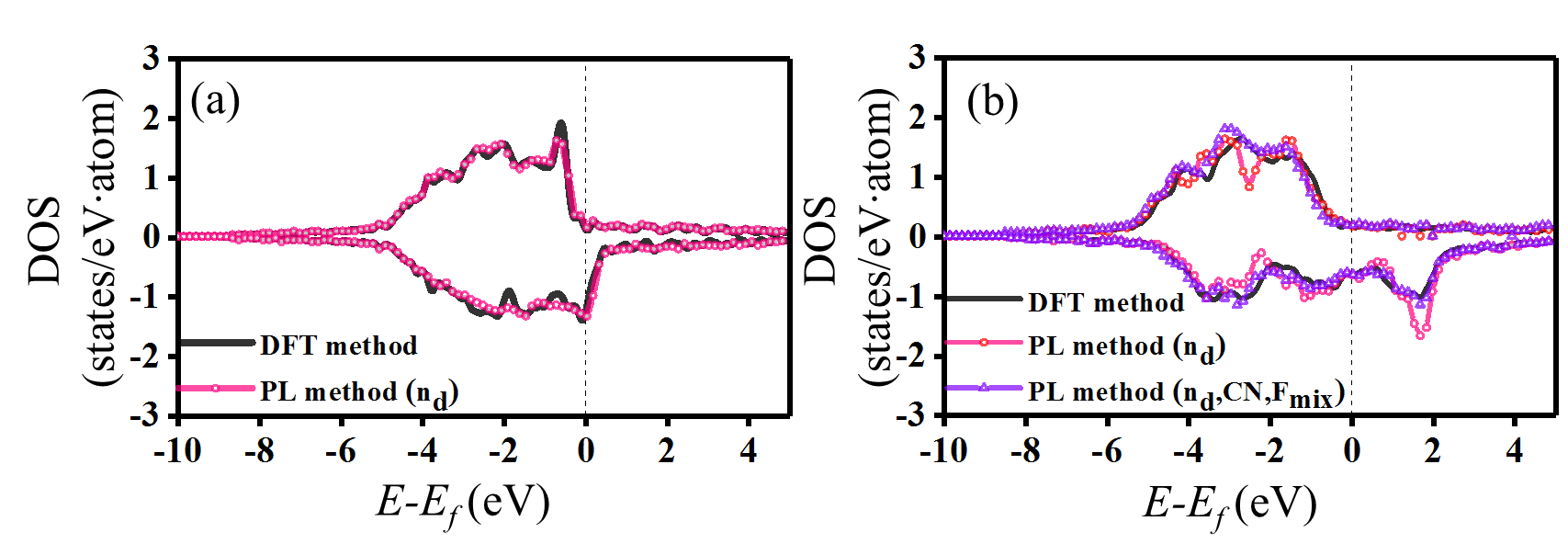}
\caption{Prediction results of the PL method in binary alloy systems. (a) DOS pattern of $\text{Cu}_{0.375}$$\text{Ni}_{0.625}$ as a test of the Cu-Ni alloy system. Its atomic structure is shown in Fig. S3. The energy range ($E$-$E_f$) is from $E$ = -10 eV to $E$ = 5 eV, and the DOS range is from 0.0 to $\pm$3.0, where the positive region is for the up-spin, and the negative region is for the down-spin. Black corresponds to the DFT method, and pink corresponds to the learning method using only one feature of nd. (b) DOS pattern of  $\text{Cu}_{0.375}$$\text{Ni}_{0.625}$ as a test of the Cu-Fe alloy system. Its atomic structure is shown in Fig. S3. Black corresponds to the DFT method, pink corresponds to the learning method using the $\text{n}_d$ feature, and violet corresponds to the learning method using all features including $\text{n}_d$, CN, $\text{F}_{mix}$.}
\label{fig:FIG. 2}
\end{figure}

In contrast to the Cu-Ni system, the Cu-Fe system was also considered because the crystal structures of Cu and Fe are different and their alloys do not exhibit a complete solid solution. This implies that features based on the atomic structures in addition to nd are required for the DOS representation. We introduced the coordination number (CN) and a mixing factor ($\text{F}_{mix}$) as features to distinguish the atomic structures. The CN was obtained by dividing the number of all bonds between two atoms by the total number of atoms in the system, where the bonds were calculated using the covalent atomic radii. $\text{F}_{mix}$ indicates the ratio of the number of different pair bonds in the alloy system to the total number of bonds. Using $\text{F}_{mix}$, one can distinguish the atomic distributions in alloy systems that have the same CN (Supplementary Fig. S2). 

To represent the DOS patterns for the test data, the coefficient $\alpha'_p$ should be determined. Since the eigenvectors obtained after PCA correspond to the PC vectors, the distributed coefficients lying on identical eigenvectors were correlated with each other. Thus, we generated linear regression lines between the $\alpha_p$ of the training data in which we focused on the linear regression line between two training data near the test composition. Then, using the features of the training and test systems, we estimated the $\alpha'_k$ contributions of $\text{n}_d$, CN, and $\text{F}_{mix}$ 
($\alpha_{p}^{n_{d}}$, $\alpha_{p}^{CN}$, $\alpha_{p}^{F_{mix}}$) for the test system using the linear regression line (Fig. 1c). We defined the set of features as $\Phi$ = \{$\text{n}_d$, CN, $\text{F}_{mix}$\}. Here, it was assumed that the three features have equal weights so that 

\begin{equation}
\alpha'_p = \sum_{\varphi \in \Phi} \beta_{\varphi} \cdot \alpha_{p}^{\varphi}
\end{equation}

where $\beta_\varphi$ is $1/3$ for all the features. A detailed description of the estimation of the coefficients is also provided in Section 3 of the Supplementary Information (Table S1 and Fig. S6).

Using these three features ($\text{n}_d$, CN, and $\text{F}_{mix}$), the DOS of $\text{Cu}_{0.375}$$\text{Ni}_{0.625}$ was predicted (Fig. 2b). The use of only nd leads to the pattern similarity of 78\%, while the use of all three features improves the pattern similarity up to 95\%. Even in the previously examined Cu-Ni system, consideration of CN and $\text{F}_{mix}$ in addition to nd can slightly improve the pattern similarity up to 96\% for $\text{Cu}_{0.375}$$\text{Ni}_{0.625}$ (Supplementary Fig. S7).

\begin{center}
\textbf{B. Application into multi-component alloy systems}
\end{center}

To extend our method to multi-component alloy systems, we also developed a method to represent the DOS patterns of ternary systems, using the example of the Cu-Ni-Pt system (Fig. 3). Fig. 3a shows a triangular composition diagram of the Cu-Ni-Pt system, where a total of 15 compositions were considered as the training set: pure 3, binary 9, and ternary 3. Similar to the previous binary cases, by determining the coefficients ($\alpha_p$) of the PCs for a ternary test composition, one can represent its DOS pattern. First, we selected three training compositions that were located closest to the test composition and calculated the distances ($d$) between the test composition and the three training compositions (Fig. 3b). Then, the $\alpha'_p$ for the DOS representation were estimated by using the features and $\alpha_p$ at the training compositions, where it was assumed on physical grounds that the DOS pattern of the test composition is represented by the highest weight for the training composition that is nearest to the test composition. 

\begin{figure}[h!]
\includegraphics[width=0.48\textwidth]{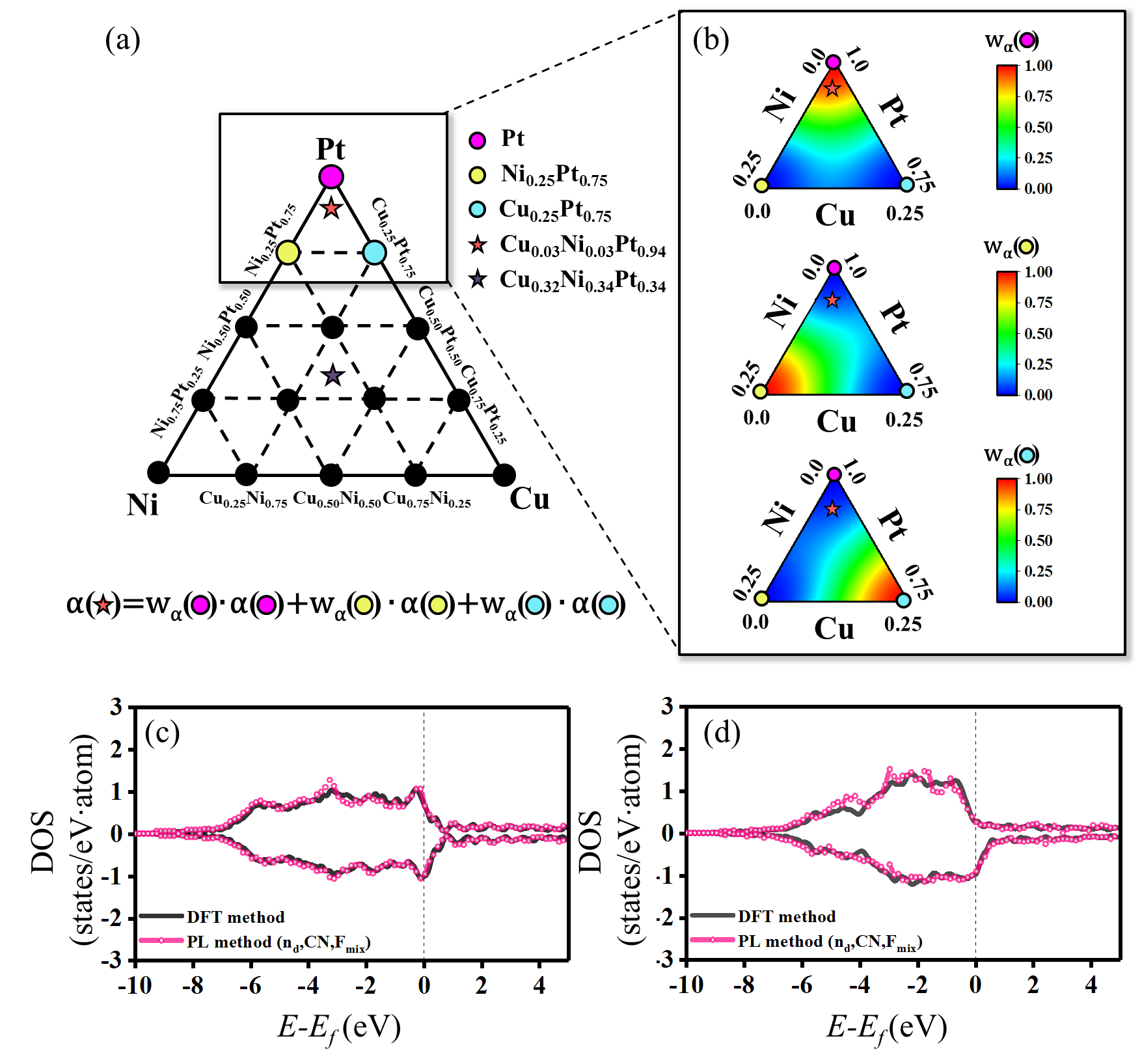}
\caption{Estimation of coefficients and prediction results of the PL method in ternary alloy systems. (a) Triangular diagram of the Cu-Ni-Pt system representing the training data (circle) and test data (star). The equation for the calculation of the PCs coefficients for the test data is shown at the bottom of the figure: the equation is based on the coefficients and their weights for training alloys that most closely match the test alloy composition. (b) Maps of the weights of the coefficients of the PC vectors for the test composition ($\text{Cu}_{0.03}$$\text{Ni}_{0.03}$$\text{Pt}_{0.94}$). The weights depends on the distance between the test composition and each training composition, and they also depend on the difference of three features ($\text{n}_d$, CN, and $\text{F}_{mix}$) between the training and test data. (c) DOS pattern of the $\text{Cu}_{0.03}$$\text{Ni}_{0.03}$$\text{Pt}_{0.94}$ test alloy. d, DOS pattern of the $\text{Cu}_{0.32}$$\text{Ni}_{0.34}$$\text{Pt}_{0.34}$ test alloy. Their atomic structures are shown in Supplementary Fig. S3. In (c) and (d), black corresponds to the DFT method, and pink corresponds to the learning method using all features including $\text{n}_d$, CN, and $\text{F}_{mix}$.}
\label{fig:FIG. 3}
\end{figure}

The basic idea for an A-B-C ternary case is similar to that of the binary case. In this case, we define the set of features as $\Phi = \{\text{n}_{d,A}, \text{n}_{d,B}, \text{n}_{d,C}, \text{CN}_{norm}, \text{F}_{mix}\}$. The number of feature values for nd depends on the number of elements in the multi-components case. Here, we also considered the differences ($d_{ij}$) in the feature values between the test and the adjacent three training compositions (Supplementary Fig. S8) as given by: 

\begin{equation}
d_{ij} = \sum_{\varphi \in \Phi} (\varphi^i - \varphi^j)^2 
\end{equation}

where $i$ and $j$ are the selected data of the A-B-C alloy system, and $\text{n}_{d,A}, \text{n}_{d,B}, \text{n}_{d,C}, \text{CN}_{norm}, \text{F}_{mix}$ are feature values corresponding to the data. Of the three material features ($\text{n}_d$, CN, $\text{F}_{mix}$), the$\text{n}_d$ and $\text{F}_{mix}$ values range from 0 to 1, whereas CN is greater than 1. To obtain units in the same range, we considered the normalized value of CN ($\text{CN}_{norm}$) by dividing the CN value by 12, which is based on the fact that the maximum CN value in the alloy system is 12 for a fcc structure. When the composition and crystal structure of a test alloy are more similar to the training data, the differences in the feature values decreases. We defined $\Omega$ as the set of the nearest three training data, $\upsilon$ as the test data, and $\upsilon'$ as the training data. To estimate the PC coefficient ($\alpha'_{k,\upsilon}$) for the test data, three weights ($w$) of the coefficients of the three training systems were calculated based on $d_{ij}$ using eq. (7):

\begin{equation}
w_{\upsilon\upsilon'}=\frac{d_{\upsilon\upsilon'}^{-1}}{\sum_{\upsilon' \in \Omega} d_{\upsilon\upsilon'}^{-1}}
\end{equation}

The range of $w_{\upsilon \upsilon'}$ is from 0 to 1. Then, the estimated PC coefficients for the test alloy were calculated by

\begin{equation}
\alpha'_{p,\upsilon} = \sum_{\upsilon' \in \Omega} w_{\upsilon \upsilon'} \cdot \alpha_{k,\upsilon'} 
\end{equation}

For the example of the Cu-Ni-Pt system shown in Fig. 3, the $\text{n}_d$, CN, $\text{F}_{mix}$ of the training and test data are summarized in Supplementary Table S2. This approach was tested for two compositions: $\text{Cu}_{0.03}$$\text{Ni}_{0.03}$$\text{Pt}_{0.94}$ (Fig. 3c) and $\text{Cu}_{0.32}$$\text{Ni}_{0.34}$$\text{Pt}_{0.44}$ (Fig. 3d), and we determined that our method obtains the pattern similarity of 96\%.

\begin{center}
\textbf{C. Application into surface structures systems}
\end{center}

By expanding the scheme that was applied to bulk structures, we studied the representation of the DOS patterns for surface structures of alloys. In particular, we used method to represent the DOS patterns of high-index surfaces based on those of low-index surfaces. Here, it is important to find a feature to define the surface structures, with which we can estimate the PC coefficients for a test surface. In Fig. 4a, a high-index surface, (211), can be regarded as a surface to connect two low-index surfaces, (011) and (111). Moreover, the step alignment of atoms on the (211) surface plane is generated by a combination of the atom alignments on the (011) and (111) surface plane. In this regard, we employed a lattice plane vector by using the miller indices, which consist of three integers, $h$, $k$, and $l$, and where the notation of the lattice plane vector is written by ($hkl$). Then, we defined the lattice plane that intercepts three points, $\vec{L}_{1}/h$, $\vec{L}_{2}/k$, and $\vec{L}_{3}/l$, where $\vec{L}_{1}$, $\vec{L}_{2}$, and $\vec{L}_{3}$ are the lattice vectors in a conventional unit cell. Therefore, we considered the inverses of the miller indices, $1/h$, $1/k$, and $1/l$, as the features regarding the surface plane orientations, and they are denoted by $h′$, $k′$, and $l′$, respectively. If one of the miller indices is zero, the feature value is set to be zero to avoid an infinity value. 

\begin{figure}[h]
\includegraphics[width=0.48\textwidth]{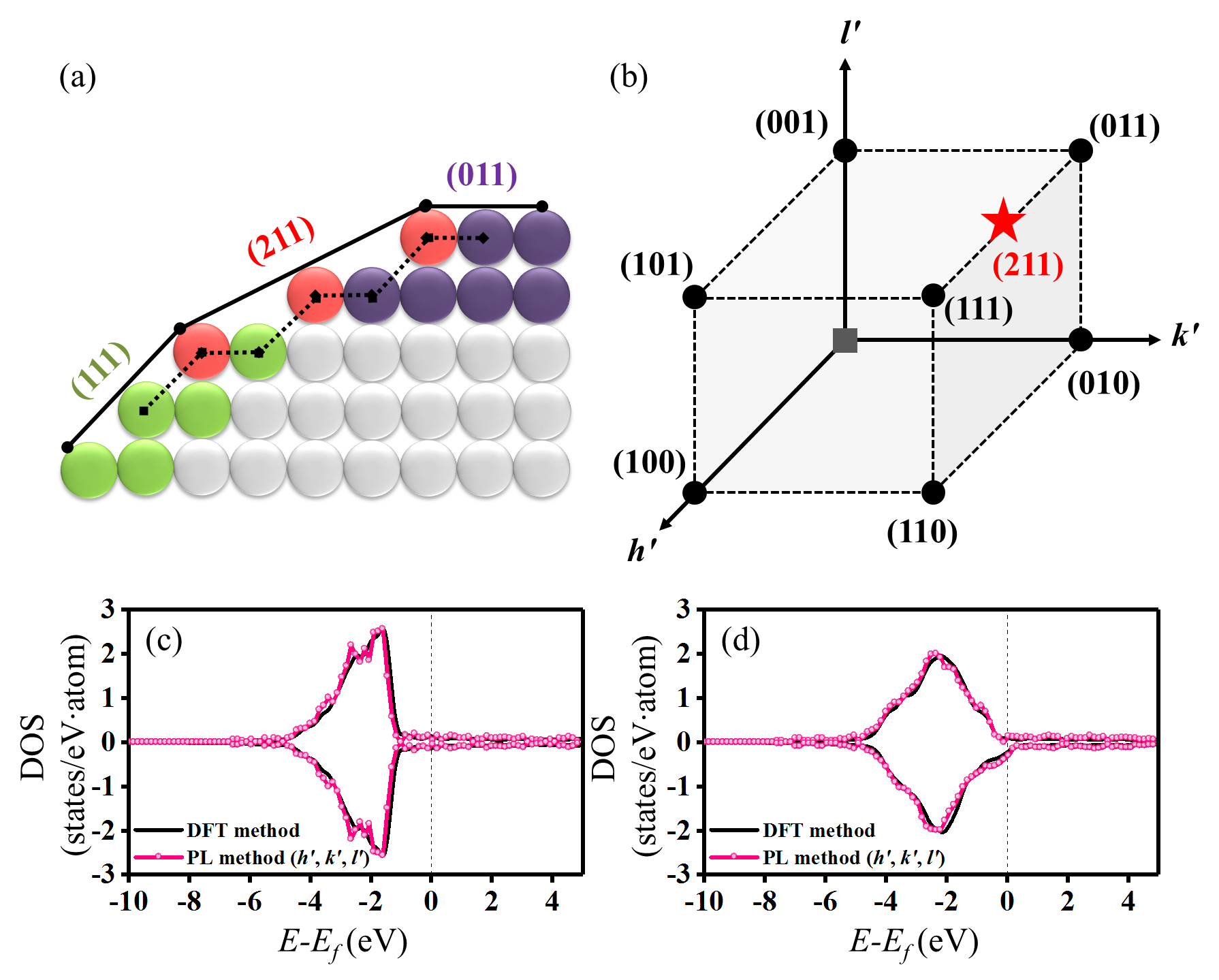}
\caption{Scheme of the PL method for the DOS representation of surface structures and the predicted results. (a) Two-dimensional cleaved lattice structure to represent the plane vectors of the (011) and (111) low-index surface and the (211) high-index surface. Here, red, green, and violet nodes represent the atoms on the surface layer for the (211), (111), and (011) plane vector, respectively. The dotted lines represent the periodicities of the (111) and (011) lattice vectors. (b) Three-dimensional cubic diagram of the lattice plane vectors in the coordinate system for the inverse of miller indices $h′$, $k′$, and $l′$ representing the training (black circle) and test (red star) data. (c) DOS pattern of the Cu (211) surface . (d) DOS pattern of the $\text{Cu}_{0.375}$$\text{Ni}_{0.625}$ (211) surface. In c and d, black corresponds to the DFT method, and pink corresponds to the learning method (use of three PCs) using the inverse value of the miller index as a feature. The atomic structures of each surfaces can be found in Supplementary Fig. S4.}
\label{fig:FIG. 4}
\end{figure}

Fig. 4b shows the three-dimensional ($h′$, $k′$, $l′$) vector space representing the inverses of miller indices for the training and test surfaces. The training samples include seven lattice plane vectors where all of the miller indices are lower than 2; \{(001), (010), (100), (011), (101), (110), (111)\}. The vectors correspond to low-index surface plane vectors. Then, after adding the origin vector, (000), and connecting all vectors of the training samples and the origin, a cubic geometrical figure can be obtained. The test sample is the vector of which the miller indices is larger than one, which corresponds to a high-index surface plane vector. Thus, the high index surface plane vectors can lie on an edge or face in the cube figure. In Fig. 4a, the alignments of atoms on the (211) surface plane is a combination of the atom alignments on the (011) and (111) surface planes. Therefore, we can estimate the DOS pattern for the (211) surface with those for the (011) and (111) surfaces. Here, for the three vectors, the miller indices $k$ and $l$ are the same, indicating that one can be distinguished only using the miller index $h$. During the predicting process of our method (Fig. 1c), we only used the $h′$ value to determine the PC coefficients, $\alpha'_{p,(211)}$, by the linearly interpolating between the two coefficients for the (011) and (111) surfaces after performing the PCA using all training DOS data.

To validate our method for surface structures, we tested (211) surfaces of the pure Cu metal (Fig. 4c) and the $\text{Cu}_{0.375}$$\text{Ni}_{0.625}$ alloy (Fig. 4d). For the Cu case, our method provided the pattern similarity of 93\% compared to DFT calculation (Fig. 4c), where we considered only three DOS data for Cu(001), (011), and (111) surfaces as the training data. When predicting the DOS pattern for the (211) surface of the $\text{Cu}_{0.375}$$\text{Ni}_{0.625}$ alloy (see Supplementary Fig. S9), we considered (001), (011), and (111) surfaces of five Cu-Ni alloys; \{Cu, $\text{Cu}_{0.75}$$\text{Ni}_{0.25}$, $\text{Cu}_{0.5}$$\text{Ni}_{0.5}$, $\text{Cu}_{0.25}$$\text{Ni}_{0.75}$, Ni\}. We first predicted the DOS patterns for (001), (011) and (111) surfaces of the $\text{Cu}_{0.375}$$\text{Ni}_{0.625}$ alloy with three features ($\text{n}_d$, CN, and $\text{F}_{mix}$), which is similar to the method used in the bulk case. Then, we performed PCA one more time for the DOS patterns for the low-index surfaces of the $\text{Cu}_{0.375}$$\text{Ni}_{0.625}$ alloy that were obtained after the first pattern learning method. Then, using the inverse value of the miller index as a feature, we predicted the DOS pattern for the (211) surface of the $\text{Cu}_{0.375}$$\text{Ni}_{0.625}$ alloy, and obtained the pattern similarity of 97\% (Fig. 4d). In DFT calculations, the slab calculations are much more time-consuming than bulk calculations. However, our method provides a similar calculation speed for both bulk and surface systems. For example, for the Cu(211) and $\text{Cu}_{0.375}$$\text{Ni}_{0.625}$(211) surface, the DFT calculation takes 2 hours on 36 cores of the CPU, while our method is still less than 5 minute even on 1 core of the CPU.

\begin{center}
\textbf{\Romannum{4}. DISCUSSION}
\end{center}

Although high-performance computing machines have been used practically thus far, we could still tackle large-scale first-principles calculations of more than hundreds atoms using the limited computing power. Regarding the computation cost, it is well-known that the DFT method scales as O($N^3$), where $N$ is the number of electrons in the system\cite{Whitfield2013}. Indeed, a similar trend was observed in this work (Fig. 5a). However, our method remarkably shows a 100 to 1,000 times higher speed than DFT and requires only 1 minute regardless of $N$, scaling as O(1).  Moreover, Fig. 5b shows the pattern similarity of our method for various binary alloy systems composed of 5 transition metals (Cu, Ni, Ru, Pd, Pt), and found that the pattern similarity is as high as 91$\sim$98\%. Our method outperforms DFT calculations in terms of the calculation speed, and it loses little information compared to the DFT electronic structures. This clearly reveals that our method will be an alternative to break the trade-off relationship between accuracy and speed, which is well known in the field of electronic structure calculations.

\begin{figure}[h]
\includegraphics[width=0.38\textwidth]{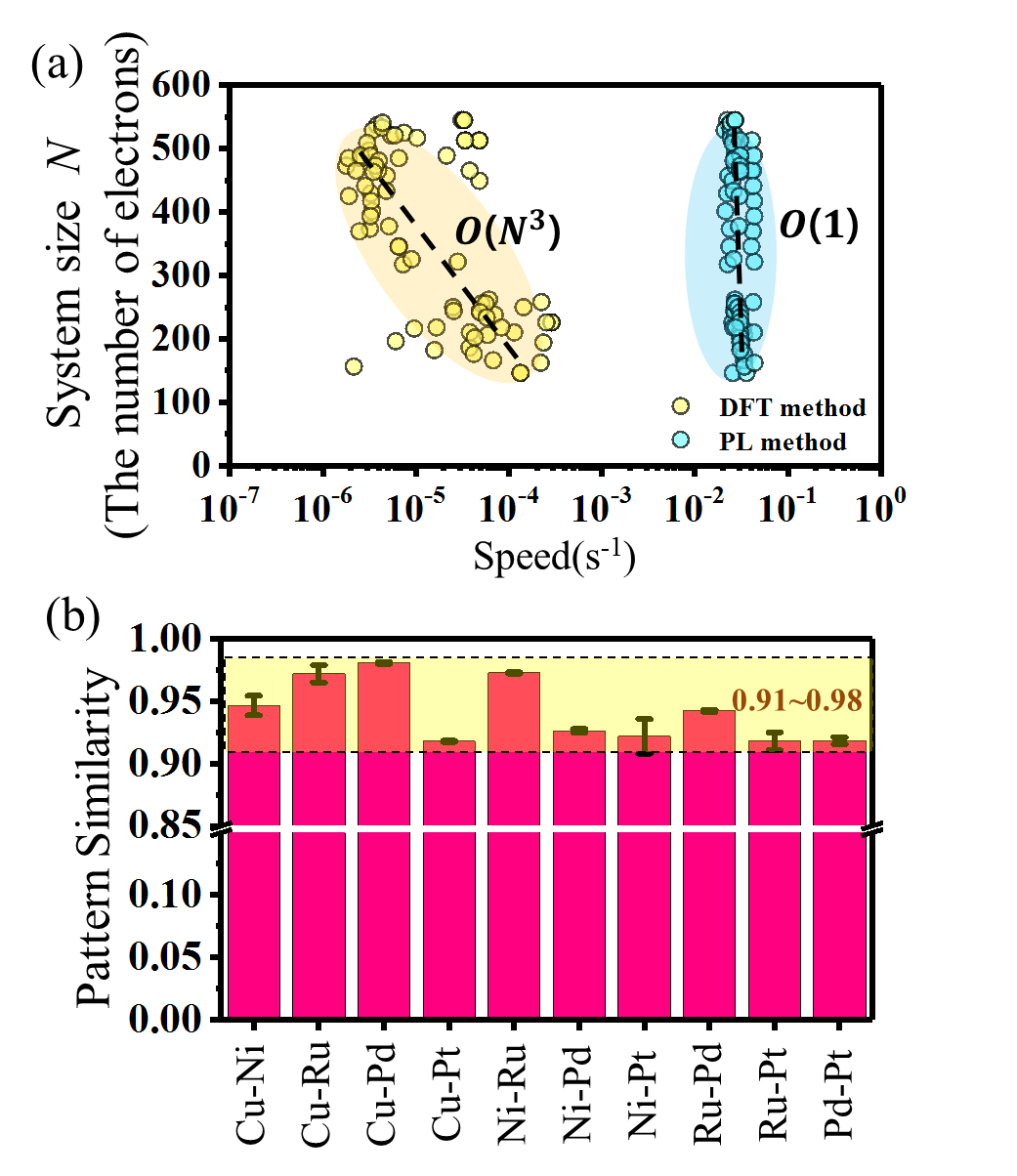}
\caption{Performances of the PL method compared with the DFT method. (a) Comparison of the calculation speeds of the learning method (cyan) and DFT (yellow) as a function of the number of electrons in the alloy systems. The learning method scales as O(1), indicating no dependence on the system size, whereas the DFT scales as O($N^3$). The calculation times for 1 core of CPU time and 80 alloy systems were considered. (b) Accuracies of the learning method for 10 test alloys in various binary alloy systems: Cu-Ni, Cu-Ru, Cu-Pd, Cu-Pt, Ni-Ru, Ni-Pd, Ni-Pt, Ru-Pd, Ru-Pt, and Pd-Pt. The yellow region is highlighted to show the pattern similarity range of our learning model (91$\sim$98\%).}
\end{figure}

One of the novelties of this work is that the electronic DOS that was originally a function of an energy level can be expressed with a simple model in a linear combination form of few (three or four) PC bases. Here, we highlight the use of only three or four PCs. Although the available number of PCs in a learning process of DOS patterns is as many as the dimension of the DOS image vectors, the number of PCs highly contributing to the representation of the DOS is few, where the contribution can be evaluated with the eigenvalue for each PC. Thus, four PCs are enough to recognize the diversity of the DOS patterns in the training data. For example, in the binary alloy systems of Fig. 2 where five training data are considered, the contribution of each PC to the representation of the DOS pattern is 32.9\% for the 1st PC, 25.7\% for the 2nd PC, 22.2\% for the 3rd PC, and 19.1\% for the 4th PC, which indicates that the contribution of the remaining PCs is very miniscule (less than 0.1\%). This clearly shows that only four PCs in the PCA are sufficient to represent DOS patterns.

The performance (calculation speed and accuracy) of our PL method is affected by the number of PCs and the grid size. Interestingly, due to the overfitting problem\cite{Dominggos2012}, the use of three PCs provides the most accurate DOS patterns in Cu-Ni alloy systems (Supplementary Fig. S10), although at least four PCs are required to fully represent the DOS patterns in training data as discussed in the above paragraph.  The grid size also affects the accuracy and calculation time of our method (Supplementary Fig. S10). The use of a higher (or finer) grid size provides a higher accuracy, even though an improvement in the accuracy for a grid with a higher density than a 100 $\times$ 100 grid ($M$ = 100) is not significant. However, for $M$ $>$ 100, the calculation time is significantly high. Thus, we should employ appropriate values with respect to the number of PCs and grid size to guarantee a high accuracy ($>$ 90\%) and low calculation time ($<$ 1 min).

\begin{center}
\textbf{\Romannum{5}. CONCLUSION}
\end{center}

To our knowledge, in this work, we presented the first machine-learning approach for calculating electronic DOS patterns (both of value and shape) with a strong accuracy and a fast speed. Moreover, our approach can handle a variety of spectrum image data of materials (X-ray photoelectron spectroscopy, X-ray diffraction, Raman spectrum, etc.). Toward an era of data-driven material design, the importance of material databases will continue to increase; however, the accumulation of data will be a serious bottleneck. In this regard, the fast generation of material databases will be a key in the future. Application of our PCA-based method into various image-type data will provide rapid and accurate prediction of various material properties in place of DFT calculations or other experimental measurements. Therefore, it is anticipated that our model will accelerate the construction of large-scale material databases as well as the design of materials in various fields such as catalysts and electronic devices\cite{Kolb2017,Hill2017}.

\begin{center}
\textbf{ACKNOWLEDGMENTS}
\end{center}

This work was supported by Creative Materials Discovery Program through the National Research Foundation of Korea (NRF-2016M3D1A1021140). We acknowledge the financial supports of the Korea Institute of Science and Technology (Grant No. 2E28000).

\begin{center}
\textbf{APPENDIX A: DATA SELECTION}
\end{center}

To represent the DOS pattern for a test alloy system with the learning model developed in this work, the relevant training data are required. The data include alloy compositions, crystal structures, and DOS patterns. In general, the more training data would provide a more accurate representation. However, since the main purpose of this work is to introduce a new scheme for obtaining DOS patterns by a machine-learning method, we used a limited data set. In a binary A-B system, we considered five data sets (two pure structures and three alloy ones). For the pure cases, we used the experimental crystal structure. For the alloy systems, the $\text{A}_{0.25}$$\text{B}_{0.75}$, $\text{A}_{0.5}$$\text{B}_{0.5}$, and $\text{A}_{0.75}$$\text{B}_{0.25}$ compositions were considered. Here, based on the thermodynamic phase diagram of the alloy system, the crystal structures of the three compositions were determined. If there exists an intermetallic phase at the alloy composition, we preferentially considered an intermetallic crystal structure (e.g.,$\text{L1}_{0}$ for $\text{Pt}_{0.5}\text{Ni}_{0.5}$, and $\text{L1}_{2}$ for $\text{Pt}_{0.25}\text{Ni}_{0.75}$ and $\text{Pt}_{0.75}\text{Ni}_{0.25}$). On the other hand, for the cases where the intermetallic phase does not exist, atomically randomly mixed structures (i.e., solid-solution phases) were considered with two crystal structures of pure A and B. Among these two structures, we selected the more stable structure as determined by the DFT calculations. The compositions and atomic structures considered in bulk and slab structures study as training and test data are described in Supplementary Fig. S3 and S4, respectively. Then, in slab structures study, the compositions of surface layers are considered as same as $\text{A}_x$$\text{B}_{1-x}$. The DOS patterns of the training structures were also obtained from the DFT calculations.

\begin{center}
\textbf{APPENDIX B: DFT CALCULATION OF ELECTRONIC STRUCTURE}
\end{center}

All electronic structure calculations were performed using the Vienna ab initio simulation package (VASP)\cite{Kresse1999,Kresse1996}. The exchange-correlation energy was described by the revised Perdew-Burke-Ernzerhof (RPBE) exchange functional\cite{Perdew1996,Hammer1999}. The electronic wave functions were expanded in the plane-wave basis set with a kinetic energy cutoff of 520 eV. The effect of the core electrons was modeled by projector augmented-wave (PAW) potentials\cite{Blochl1994}. The Brillouin zone was sampled using a Monkhorst-Pack $k$-point mesh, and the $k$-point sampling was set to $8\times8\times8$ for bulk structures and $4\times4\times1$ for slab structures. The bulk crystal structures were modeled using a $2\times2\times2$ supercell (e.g., fcc: 32 atoms and bcc: 16 atoms), and the slab crystal structures were simulated periodically with four layer cells. In slab structures calculations, a large vacuum spacing $\>$15 $\AA$ was used to prevent inter-slab interactions, and the top most surface layer and sub-surface layer of the computational cell were geometrically relaxed such that the maximum force on each atom was less than 0.05 eV ${\AA}^{-1}$. Their DOS patterns were obtained after a geometry optimization process. We focused on the local DOS of the $d$ orbitals in metals for simplicity, and every DOS normalized by the number of atoms in a periodic system was described as DOS = $f (E-E_f)$, where $E-E_f$ is the relative energy shift from the Fermi level ($E_f$). In addition, during the DFT calculations, we turned on the spin polarization effect to consider the magnetic properties of the metals. In representing the DOS patterns via our learning model, we applied our model separately for the up spin and the down spin. 

\begin{center}
\textbf{APPENDIX C: PATTERN SIMILARITY CALCULATION}
\end{center}

The pattern similarity of our learning model was calculated through a comparison with the DFT results, in which the $l^2$-norm was used. The pattern similarity $\sigma$ is defined as follows:

\begin{equation}
\sigma=1-\frac{\sqrt{\sum_{m=1}^{M} |\rho'(E_m)-\rho(E_m)|^2}}{\sqrt{\sum_{m=1}^{M} |\rho(E_m )|^2}} 
\end{equation}

where $\rho'$ and $\rho$ are the DOS patterns obtained by our learning method and calculated by the DFT method, respectively. When $\sigma$ is closer to 1, our method becomes more accurate.


\begin{references}

\bibitem{Martin} R. M. Martin, Electronic Structure: Basic Theory and Practical Methods (Cambridge Univ. Press, 2004).

\bibitem{Norskov2009} J. K. N$\o$rskov, T. Bligaard, J. Rossmeisl, and C. H. Christensen, Towards the computational design of solid catalysts, Nat. Chem. {\bf1}, 121 (2009).

\bibitem{Seo2014} D. Seo, H. Shin, K. Kang, H. Kim, and S. S. Han, First-Principles Design of Hydrogen Dissociation Catalysts Based on Isoelectronic Metal Solid Solutions, J. Phys. Chem. Lett. {\bf5}, 1819 (2014).

\bibitem{Ma2015} X. Ma, Z. Li, L. E. K. Achenie, and H. Xin, Machine-Learning-Augmented Chemisorption Model for $\text{CO}_2$ Electroreduction Catalyst Screening, J. Phys. Chem. Lett. {\bf6}, 3528 (2015).

\bibitem{Ratcliff2017} L. E. Ratcliff, S. Mohr, G. Huhs, T. Deutsch, M. Masella, and L. Genovese, Challenges in large scale quantum mechanical calculations, WIREs Comput. Mol. Sci. {\bf7}, 1 (2017).

\bibitem{Galli1996} G. Galli, Quantum Molecular Dynamics Simulations, Curr. Opin. Solid State Mater. Sci. {\bf1}, 864 (1996).

\bibitem{Saad2010} Y. Saad, J. R. Chelikowsky, and S. M. Shontz, Numerical Methods for Electronic Structure Calculations of Materials, SIAM Rev. {\bf52}, 3 (2010).

\bibitem{Goedecker1999} S. Goedecker, Linear Scaling Electronic Structure Methods, Rev. Mod. Phys. {\bf71}, 1085 (1999).

\bibitem{Ordej1998} P. Ordej, Order-$\textit{N}$ tight-binding methods for electronic-structure and molecular dynamics, Comput. Mater. Sci. {\bf12}, 157 (1998).

\bibitem{Lecun2015} Y. Lecun, Y.  Bengio, and G. Hinton, Deep learning, Nature {\bf521}, 436 (2015).

\bibitem{Jordan2015} M. I. Jordan, and T. M. Mitchell, Machine learning: Trends, perspectives, and prospects, Science  {\bf349}, 255 (2015).

\bibitem{Carleo2017} G.  Carleo, and M. Troyer, Solving the quantum many-body problem with artificial neural networks, Science {\bf606}, 602 (2017).

\bibitem{Biamonte2017} J. Biamonte, P. Wittek, N. Pancotti, P. Rebentrost, N. Wiebe, and S. Lloyd, Quantum machine learning, Nature  {\bf549}, 195 (2017).

\bibitem{Carrasquilla2017} J. Carrasquilla, and R. G. Melko, Machine learning phases of matter, Nat. Phys. {\bf13}, 431 (2017).

\bibitem{Nieuwenburg2017} E. P. L. V. Nieuwenburg, Y. Liu, and S. D. Huber, Learning phase transitions by confusion. Nat. Phys.  {\bf13}, 435 (2017).

\bibitem{Snyder2012} J. C. Snyder, M. Rupp, K. Hansen, K. Mu, and K. Burke, Finding Density Functionals with Machine Learning. Phys. Rev. Lett.  {\bf108}, 253002 (2012).

\bibitem{Ghiringhelli2015} L. M. Ghiringhelli, J. Vybiral, S. V. Levchenko, C. Draxl, and M. Scheffler, Big Data of Materials Science: Critical Role of the Descriptor, Phys. Rev. Lett. {\bf114}, 105503  (2015).

\bibitem{Brockherde2017} F. Brockherde, L. Vogt, L. Li, M. E. Tuckerman, K. Burke, and K. Muller, Bypassing the Kohn-Sham equations with machine learning, Nat. Commun. {\bf8}, 872 (2017).

\bibitem{Arsenault2014} L. Arsenault, A. Lopez-bezanilla, and A. J. Millis, Machine learning for many-body physics: The case of the Anderson impurity model, Phys. Rev. B  {\bf90}, 155136  (2014).

\bibitem{Schutt2014} K. T. Schutt, H. Glawe, F. Brockherde, A. Sanna, and E. K. U. Gross, How to represent crystal structures for machine learning: Towards fast prediction of electronic properties, {\bf89}, 205118  (2014).

\bibitem{Takigawa2016} I. Takigawa, K. Shimizu, K. Tsuda, and S. Takakusagi, Machine-learning prediction of the d-band center for metals and bimetals, RSC Adv.  {\bf6}, 52587 (2016).

\bibitem{Mueller2016} T. Mueller, A. G. Kusne, and R. Ramprasad, Machine Learning in Materials Science: Recent Progress and Emerging Applications, Rev. in Comput. Chem.  {\bf29}, 186 (2016).

\bibitem{Hastie} T. Hastie, R. Tibshirani, and J. Friedman, The Elements of Statistical Learning (Springer, 2009)

\bibitem{Ramani2017} V. Ramani, X. Deng, R. Qiu, K. L. Gunderson, F. J. Steemers, C. M. Disteche, W. S. Noble, Z. Duan, and J. Shendure, Massively multiplex single-cel Hi-C, Nat. Meth. {\bf14}, 263 (2017).

\bibitem{Whitfield2013} J. D. Whitfield, P. J. Love, and A. Aspure-Guzik, Computational complexity in electronic structure, Phys. Chem. Chem. Phys.  {\bf15}, 397 (2013).

\bibitem{Dominggos2012} P. A. Dominggos, Few Useful Things to Know About Machine Learning, Communications of the ACM {\bf55}, 78 (2012).

\bibitem{Kolb2017} B. Kolb, L. C. Lentz, and A. M. Kolpak, Discovering charge density functionals and structure-property relationships with PROPhet: A general framework for coupling machine learning and first- principles methods, Sci. Rep. {\bf7}, 1 (2017).

\bibitem{Hill2017} J. Hill, G. Mulholland, K. Persson, R. Seshadri, C. Wolverton, and B. Meredig, Materials science with large-scale data and informatics: Unlocking new opportunities, MRS Bulletin,  {\bf41}, 399 (2017).

\bibitem{Kresse1999} G. Kresse, and D. Joubert, From Ultrasoft Pseudopotentials to the Projector Augmented-Wave Method, Phys. Rev. B {\bf59}, 1758 (1999)

\bibitem{Kresse1996} G. Kresse, and J. Furthmiiller, Efficiency of Ab-Initio Total Energy Calculations for Metals and Semiconductors Using a Plane-Wave Basis Set, Comput. Mater. Sci.  {\bf6}, 15 (1996).

\bibitem{Perdew1996} J. P. Perdew, K. Burke, and M. Ernzerhof, Generalized Gradient Approximation Made Simple, Phys. Rev. Lett. {\bf3}, 3865 (1996).

\bibitem{Hammer1999} B. Hammer, L. B. Hansen, and J. K. N$\o$rskov, Improved Adsorption Energetics within Density-Functional Theory Using Revised Perdew-Burke-Ernzerhof Functionals, Phys. Rev. B {\bf59}, 7413 (1999).

\bibitem{Blochl1994} P. E. Blochl, Projector Augmented-wave, Phys. Rev. B {\bf50}, 17953 (1994).

\end{references}
\end{document}